\documentclass{aip-cp}

\usepackage[numbers, compress]{natbib}
\usepackage{rotating}
\usepackage{graphicx}
\usepackage[utf8]{inputenc}

\newcommand{\dd}{ {\textrm d}}
\begin{document}
\title{Application of the Non-extensive Statistical Approach to High Energy Particle Collisions}

\author[aff1,aff2]{G\'abor B\'ir\'o\corref{cor1}}
\author[aff1]{Gergely G\'abor Barnaf\"oldi}
\author[aff1]{Tam\'as S\'andor Bir\'o}
\author[aff1,aff3]{K\'aroly \"Urm\"ossy}

\affil[aff1]{Wigner Research Center for Physics of the HAS\\29--33 Konkoly--Thege Mikl\'os Str. H-1121 Budapest, Hungary}
\affil[aff2]{E\"otv\"os Lor\'and University\\1/A P\'azm\'any P\'eter S\'et\'any, H-1117  Budapest, Hungary}
\affil[aff3]{Shandong University, Jinan, China}
\corresp[cor1]{Corresponding author: biro.gabor@wigner.mta.hu}

\maketitle

\begin{abstract}
In high-energy collisions the number of the created particles is far less than the thermodynamic limit, especially in small colliding systems (e.g. proton-proton). Therefore final-state effects and fluctuations in the one-particle energy distribution are appreciable. As a consequence the characterization of identified hadron spectra with the Boltzmann\,--\,Gibbs thermodynamical approach is insufficient~\cite{artic:tsbentr}.
Instead particle spectra measured in high-energy collisions can be described very well with Tsallis\,--\,Pareto distributions, derived from non-extensive thermodynamics~\cite{artic:tsbphysica, artic:tsbeurphys}. Using the Tsallis q-entropy formula, a generalization of the Boltzmann\,--\,Gibbs entropy, we interpret the microscopical physics by analysing the Tsallis $q$ and $T$ parameters. 
In this paper we give a quick overview on these parameters, analyzing identified hadron spectra from recent years in a wide center of mass energy range. We demonstrate that the fitted Tsallis-parameters show dependency on this energy and on the particle species. Our findings are described well by a QCD inspired evolution ansatz. 
\end{abstract}
\section{INTRODUCTION}

High-energy particle accelerators have already reached the energy range where the matter of the early Universe was formed. This so called Quark Gluon Plasma (QGP) existed shortly after the Big Bang. Its properties are measured in ultrarelativistic heavy-ion collisions, where the energy density is high enough to form QGP for a short, $\mathcal{O}$(fm/$c$) time. Due to the nature of the strong interaction there is no method to observe directly this matter, only signatures from the final state allow us to draw conclusions, after the cooling and expansion of this extreme matter. Today, it is still a question, how color degrees of freedom neutralize and how quarks and gluons confine into hadrons. The {\sl hadronization} occurs during a very short time, therefore we have very limited information about it. We also do not have a well established and proofed theory of hadronization. 

Recent, complex detector systems, like ALICE at the Large Hadron Collider (CERN LHC) or STAR and PHENIX at Relativistic Heavy Ion Collider (BNL RHIC) are able to measure with high accuracy the final state particles that reach the detectors. The hadron spectra, measured in high-energy collisions, are one of the most fundamental characteristics of the events and involve both the effects of microscopical processes and collectivity in high-energy collisions. As it was found, these properties occur not only in heavy-ion collisions, but even for the case of small colliding systems like proton-proton or electron-positron collisions~\cite{artic:flowing0, artic:flowing1, artic:flowing2, artic:flowing3, artic:gbiro}. A detailed analysis of hadron spectra in terms of collective parameters is a key task for understanding hadronization.

Earlier studies show that hadron spectra can be described very well with Tsallis\,--\,Pareto-like distributions~\cite{artic:tsbeurphys, artic:gbiro, artic:tsbeurphys2,  artic:cleymansphyslett, artic:cleymansjphys, artic:wilk1, artic:wilk2, artic:wilk3}, both at low-, and high transverse momenta. This distribution has been derived in the framework of non-extensive thermodynamics. The $q$ and $T$ parameters carry important physical information of the observed system. To test this concept in more details we perform a systematic analysis of recent data, accumulating information on processes during the collisions.
 
In this paper we put emphasis on the investigation of the center-of-mass energy dependence of the Tsallis parameters $q$ and $T$, assuming a QCD inspired double-log scaling of these parameter. We use the unit scale $\hbar=c=k_B=1$.


\section{ANALYSIS DESCRIPTION}

Identified particle spectra measured in the midrapidity region in high-energy proton-proton collisions can be divided into  low- (soft) and high (hard) transverse momentum parts. While the soft, non-perturbative regime can more or less be approximated by an exponential Boltzmann\,--\,Gibbs distribution, the hard part is fitted extremely well with power-law tailed distributions, in agreement with perturbative-QCD calculations. Previous studies showed that both regions can be fitted simultaneously with a single Tsallis\,--\,Pareto distribution~\cite{artic:cleymansphyslett,artic:cleymansjphys,artic:wilk1,artic:wilk2,artic:wilk3,artic:tsorig,artic:wilk4}:
\begin{equation}
f(m_T)=A\cdot\left(1+\frac{q-1}{T} (m_T-m) \right)^{-\frac{1}{q-1}} \ \ \ ,
\label{eq:TS}
\end{equation}
where $q$ is the so called non-extensivity parameter, $T$ is a temperature-like parameter and $m_T=\sqrt{p_T^2+m^2}$ is the transverse mass including the rest mass, $m$ of the given identified hadron species. 

The above Tsallis\,--\,Pareto distribution can be regarded as a generalization of the usual Boltzmann\,--\,Gibbs distribution, which provides the opportunity to model spectra from proton-proton and heavy-ion collisions using two parameters~\cite{book:ts}. Moreover, it can be derived from the Tsallis entropy formula, where strong coupling is assumed between degrees of freedom due to the non-additivity of the standard Boltzmannian entropy. A similar behaviour is expected in strongly interacting matter if the system size is finite and fluctuations matter. Analysis of the identified hadron distributions are expected to provide information on the systems both microscopically and in general. The physical meaning of the $q$ and $T$ parameters are discussed in Refs.~\cite{artic:tsbentr, artic:tsbphysica, artic:tsbeurphys, artic:tsbeurphys2}:
\begin{equation}
 q=1-\frac{1}{C}+\frac{\Delta \beta^2}{\left< \beta\right>^2} \  \ \ \ \ \textrm{and} \ \ \ \ \  1/T=\left<\beta\right>=\left<S'(E)\right>,  
\end{equation}
where $C=\dd E/\dd T$ total heat capacity and $\beta = S'(E)$. $\Delta\beta^2$ is the variance of $\beta$. 

Our aim here is to explore the center-of-mass energy evolution of the parameters $q$ and $T$. We would like to probe the QCD-inspired double-log scaling, $\sim \log (\log \sqrt{s})$ observed in fits of electron-positron data~\cite{artic:flowing0}. For our analysis we use identified hadron spectra datasets measured in proton-proton collisions from recent years~\cite{artic:62gevphenix, artic:200gevphenix, artic:200gevstar,artic:097tevalice, artic:09tevalice,artic:276tevalice,artic:276tevpi0alice,artic:7tevalice}. The numerical fits of the various datasets were made utilizing the CERN Root program, version 6.06/00\footnote{The CERN Root analysis software is available at {\tt https://root.cern.ch/}}. Please note that it is difficult to compare the fit-parameter values for all existing data, since kinematical ranges vary and multiplicity-classes are not defined evenly. This always generates some uncertainty to the fits. Moreover, the fit results by the Root program are sensitive to the input parameters, therefore we perform the fit procedure in multiple steps:
\begin{description}
  \item{1.} fit of the high-$p_T$ part by fix $T$ and changing $q$;
  \item{2.} fit of the low-$p_T$ part by fix $q$ and changing $T$;
  \item{3.} fit of the whole $p_T$ range with free change of both parameters, starting with the above obtained $q$ and $T$.
\end{description}
Here, we set the 'low-$p_T$' as lower than $p_T< 4$ GeV and the 'high-$p_T$' part as $p_T\geq 2$ GeV, respectively. 
In the function, defined in Eq.~(\ref{eq:TS}), the parameter $T$ sets the scale of the exponential slope. In fact, for $q\longrightarrow 1$ one obtains $f_1(m_T)=Ae^{-(m_T-m)/T}$. The parameter $q$ is linked to the 'power-law like' tail at high-$p_T$. 
This procedure was tested by comparing $\chi^2/NDF$ values. 

\begin{figure}[p]
  \centerline{\includegraphics[width=\textwidth]{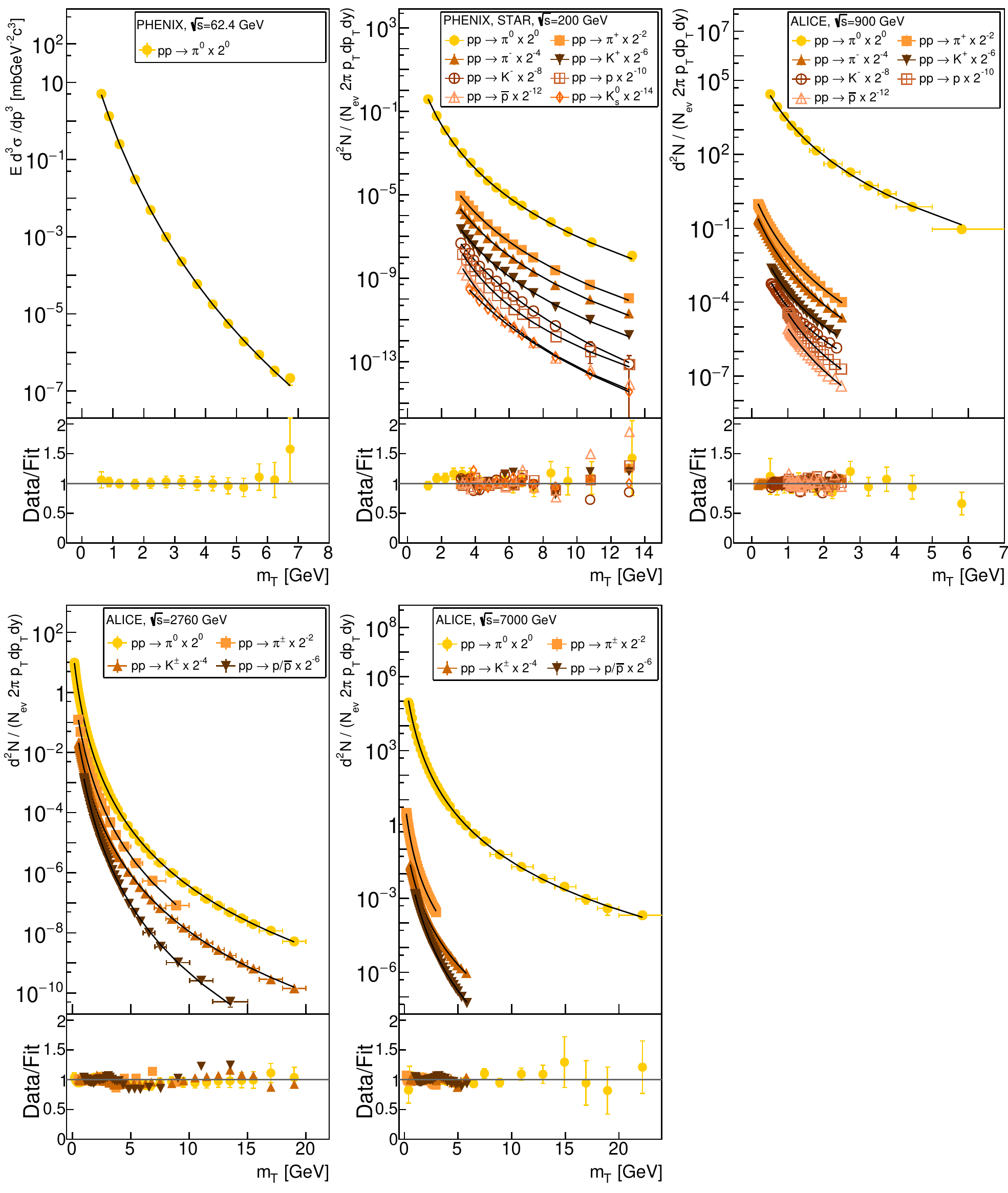}}
  \caption{\textit{Upper panels} are for the identified hadron spectra as the function of the transverse mass, $m_T$. Data were plotted, measured   by the STAR~\citep{artic:200gevstar}, PHENIX~\citep{artic:62gevphenix, artic:200gevphenix} and ALICE~\citep{artic:097tevalice,artic:09tevalice,artic:276tevalice,artic:276tevpi0alice,artic:7tevalice} at different center-of-mass energies from 62.4 GeV $\leq \sqrt{s} \leq 7$ TeV. Experimental data points are in comparison with the fitted Tsallis\,--\,Pareto functions is indicated as solid lines. The $m_T$ spectra were scaled by constant factors ($2^n$) for the better visibility as indicated on the graphs. \textit{Lower panels} present the Data/Fit ratio plots including the estimated fit errors.}
\label{fig:dataperfit}
\end{figure}

The investigated spectra together with the fitted Tsallis\,--\,Pareto function are shown in Fig.~\ref{fig:dataperfit}. {\sl Upper panels} of the plots present  the fits of experimental data measured in proton-proton collisions at $\sqrt{s}=62.4$ GeV, 200 GeV, 900 GeV, 2.76 TeV, and 7 TeV center-of-mass energies. We considered various charged and charged-averaged hadron species, $\pi^{\pm}$, $\pi^{0}$, $K^{\pm}$, $K_s^0$, $p$, and $\bar{p}$, but only the pions were measured at each energy. Identified hadron $m_T$ spectra are scaled by constant factors ($2^n$) for better visibility, as indicated in the panels.

In the {\sl lower panels} 'Data/Fit' plots are presented for each case. One can observe how well the distribution~(\ref{eq:TS}) describes the yields in the whole 62.4 GeV $\leq \sqrt{s} \leq 7$ TeV center-of-mass energy range in the $m_T \lesssim 20$ GeV region. Within the mid $m_T$-regime the overlap with data amounts to $20\%-40\%$, while at the highest $m_T$ values or for heavier hadrons the deviance is sometimes larger.

\section{FITTED PARAMETERS}

\begin{figure}[!h]
  \centerline{\includegraphics[width=0.65\textwidth]{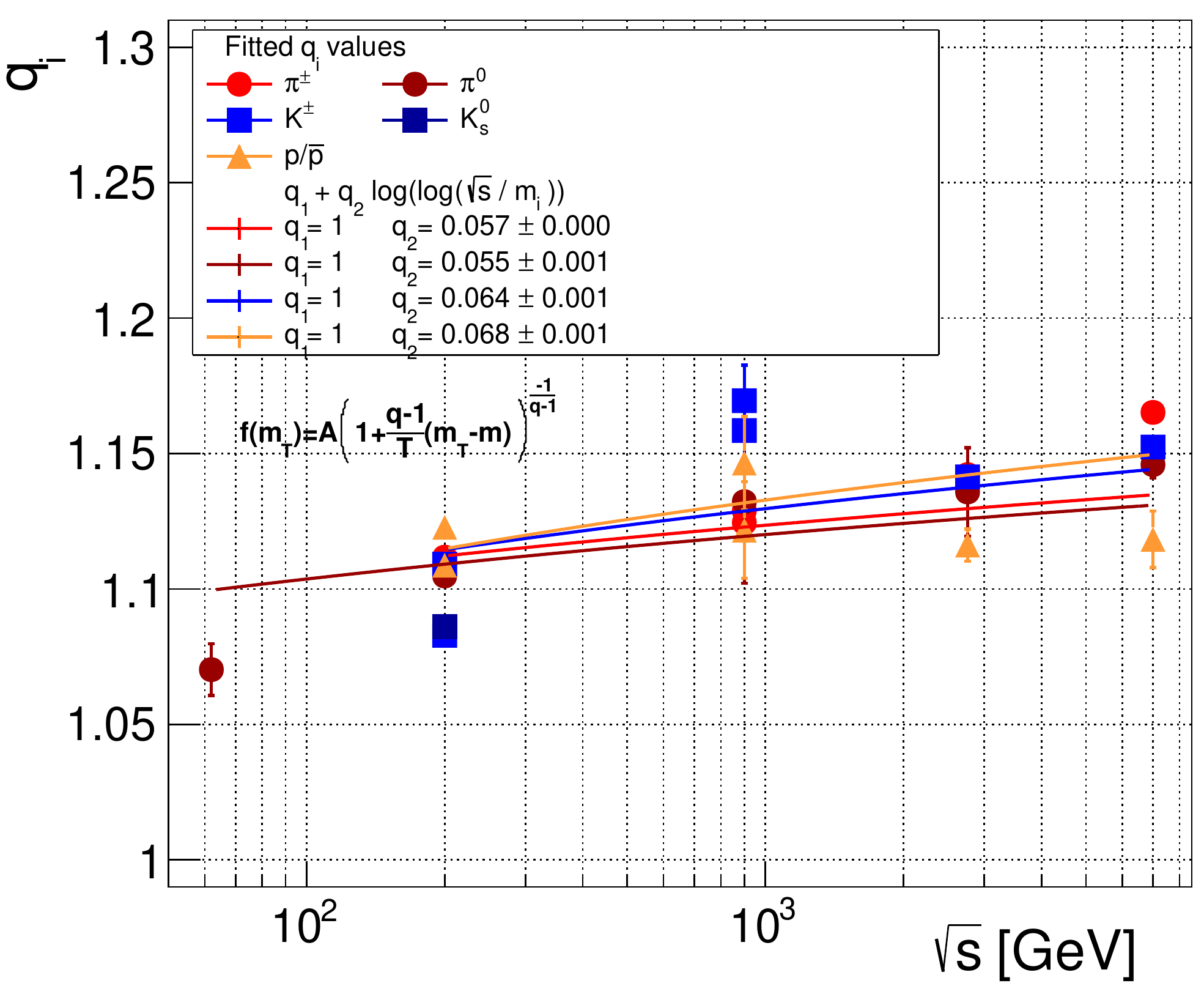}}
  \caption{The fitted $q_i$ as a function of $\sqrt{s}$ for hadron species, $i$ marked as points. Only the species of the particles are marked. The solid color lines are fitted to the pion, kaon, and (anti)proton points.}
\label{fig:q}
\end{figure}

\begin{figure}[!h]
  \centerline{\includegraphics[width=0.65\textwidth]{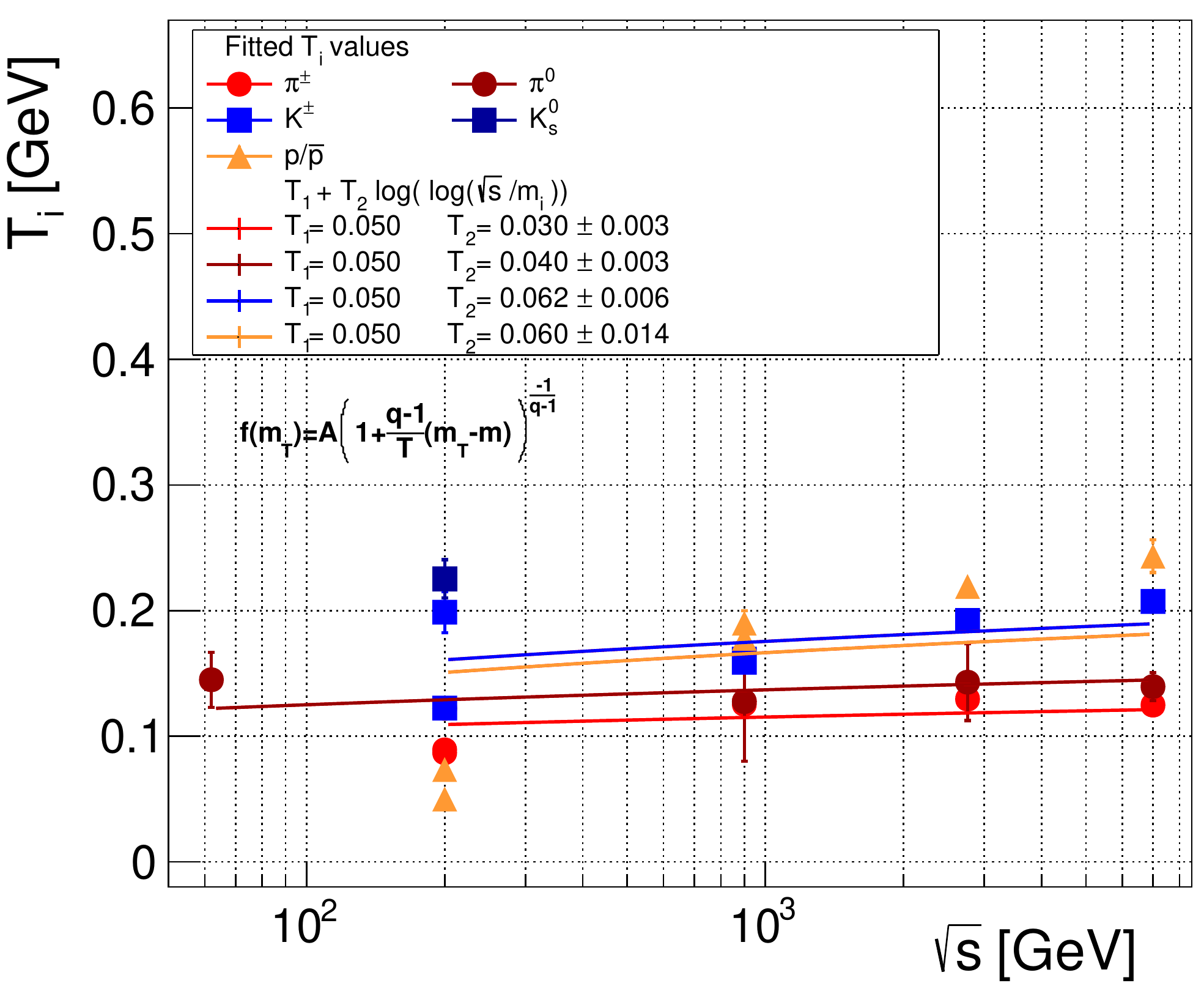}}
  \caption{The fitted $T_i$ as a function of $\sqrt{s}$ for hadron species, $i$ marked as points. Only the species of the particles is marked. The solid color lines are fitted to the pion, kaon, and (anti)proton points.}
\label{fig:T}
\end{figure}

In order to determine the center-of-mass energy dependence, in this section we review the $\sqrt{s}$ evolution of the fitted $q_i$ and $T_i$ parameters for each hadron species, $i \in ~\left\{ \pi^{\pm}, \pi^{0}, K^{\pm}, K_s^0, p, \textrm{and} \bar{p} \right\} $. According to the formula~(\ref{eq:TS}), the parameters of the Tsallis\,--\,Pareto distribution are plotted in Fig.~\ref{fig:q} and~\ref{fig:T} as a function of $\sqrt{s}$. In this sense we assumed an energy-evolution for each hadron type $i$ as follows~\cite{artic:urmossyqevol}:
\begin{equation}
\label{eq:q-evol}
q_i(\sqrt{s})=q_{1i}+ q_{2i} \log\left( \log\left( \sqrt(s)/m_i\right)\right),
\end{equation}
\begin{equation}
\label{eq:T-evol}
T_i(\sqrt{s})=T_{1i}+ T_{2i} \log\left( \log\left( \sqrt(s)/m_i\right)\right).
\end{equation}
We use the mass, $m_i$ for the identified hadrons, $i$, in order to get the physically relevant energy scale. We use the leading parameter, $q_{1i}=1.0$, representing the Boltzmann\,--\,Gibbs case at the low-energy limit and the evolution is followed only in the parameter $q_{2i}$. The temperature like parameter in the formula~(\ref{eq:T-evol}) also uses a fix $T_{1i}\approx 50$ MeV, suggested by the model in Ref~\cite{artic:tsbeurphys} for all hadron species. Here, $T_{2i}$ is the parameter, which reflects the evolution.   

On Fig.~\ref{fig:q} the fitted $q_i$ values are plotted for each $\sqrt{s}$, for the given identified hadron spectra summarized in Table~\ref{tab:qT}. One can see on the graphs, that the $q_i(\sqrt{s})$ values are close to each other and all curves slightly increase with $\sqrt{s}$, following nicely the formula~(\ref{eq:q-evol}). For pions and kaons the increase is very similar and their evolutions are alike. However, the increase seems to be larger for the kaons. 
In contrast to the mesonic components it seems that for the protons (the baryonic component) $q_{2p}$ has the strongest increase. 

\begin{table}[!h]
\centering
\label{tab:qT}
\begin{tabular}{lccccc}
\hline
Hadron, $i$ & $m_i$ & $q_{1i}$ & $q_{2i}$ & $T_{1i}$ & $T_{2i}$ \\
\hline
\hline
$\pi^{\pm}$ & 140.0 MeV & 1.0 & $0.057\pm 0.001 $ & 50 MeV & $30.0\pm 3.0 $ MeV \\
$\pi^{0}$ & 135.0 MeV &  1.0 &  $0.055\pm 0.001 $ & 50 MeV & $40.0\pm 3.0 $ MeV  \\
$K^{\pm} $ & 493.0 MeV & 1.0 &  $0.064\pm 0.001 $ & 50 MeV & $62.0\pm 6.0 $ MeV  \\
$p(\bar{p})$ & 938.0 MeV & 1.0 & $0.068\pm 0.001 $ & 50 MeV & $60.0\pm 14. $ MeV \\ 
\hline
\end{tabular}
\caption{The $\sqrt{s}$-evolution of the parameters of the fitted Tsallis\,--\,Pareto distributions for hadrons, $i \in$ $\pi^{\pm}$, $\pi^{0}$, $K^{\pm}$, $p$, and $\bar{p}$ in the $62.4$ GeV $\leq \sqrt{s} \leq 7$ TeV c.m. energy range.}
\end{table}

Figure~\ref{fig:T} presents the evolution of the parameter $T$ with the center of mass energy. We applied a $\sim \log(\log(\sqrt{s}))$-like evolution on the Fig.~\ref{fig:T} using formula~(\ref{eq:T-evol}) with the evolution parameters listed in Table~\ref{tab:qT}. Here the energy-evolution of the parameter $T(\sqrt{s})$ also shows an increasing trend with the mass of the hadron species. The obtained parameter values clearly show a mass (or quantum number conservation) effect: higher the mass, $m_i$, the parameter $T_{2i}$ is getting larger. On the other hand, one can observe, that the parameters $T_i(\sqrt{s})$ stay almost constant in the whole range of the investigated center-of-mass energies, $62.4$ GeV $\leq \sqrt{s} \leq 7$ TeV.

\begin{figure}[!h]
	\centering
	\begin{minipage}[c]{0.45\textwidth}
		\centering
	    \includegraphics[width=1\textwidth]{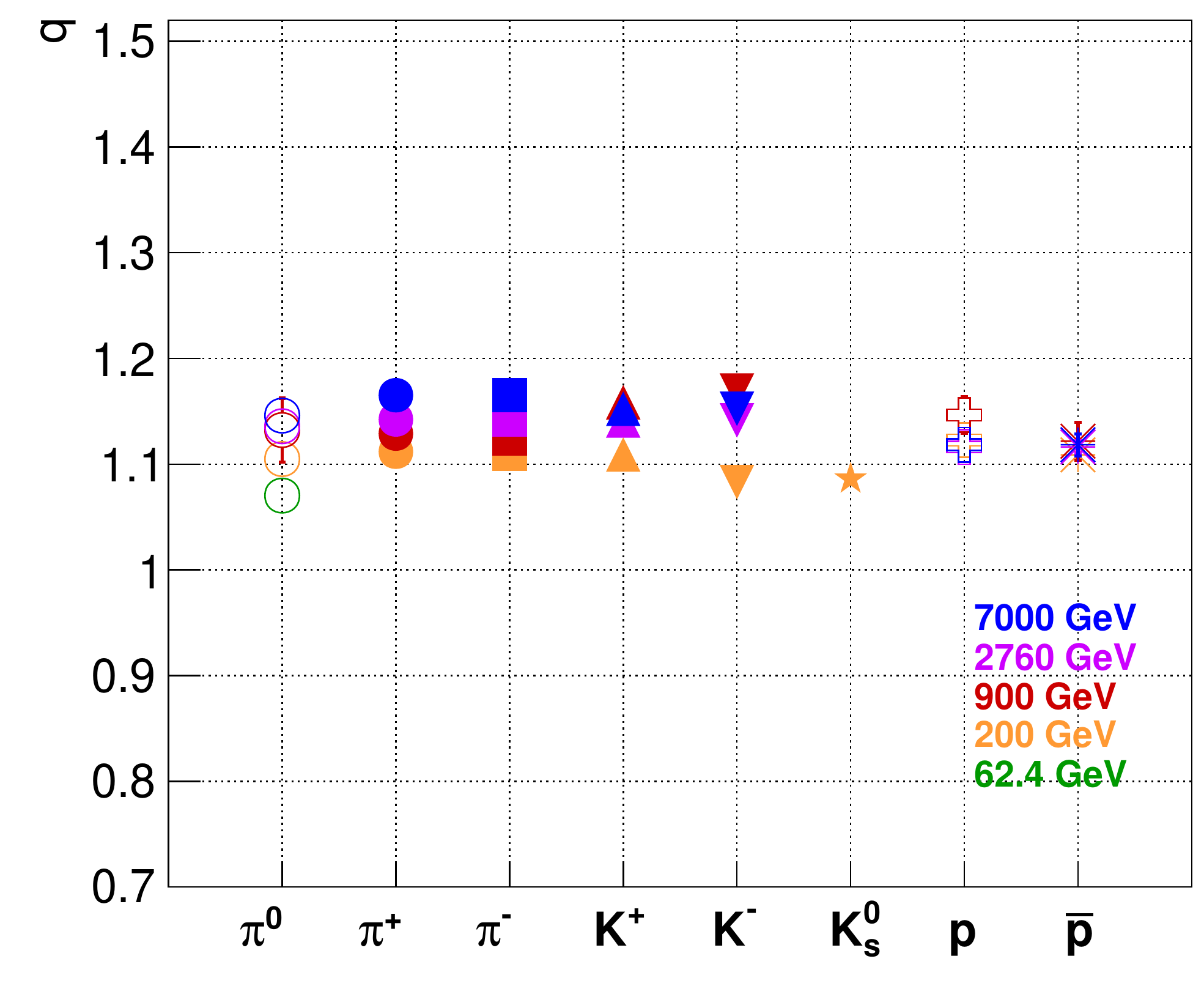}
	  \caption{}
    \end{minipage}
  	\begin{minipage}[c]{0.45\textwidth}
  		\centering
	    \includegraphics[width=1\textwidth]{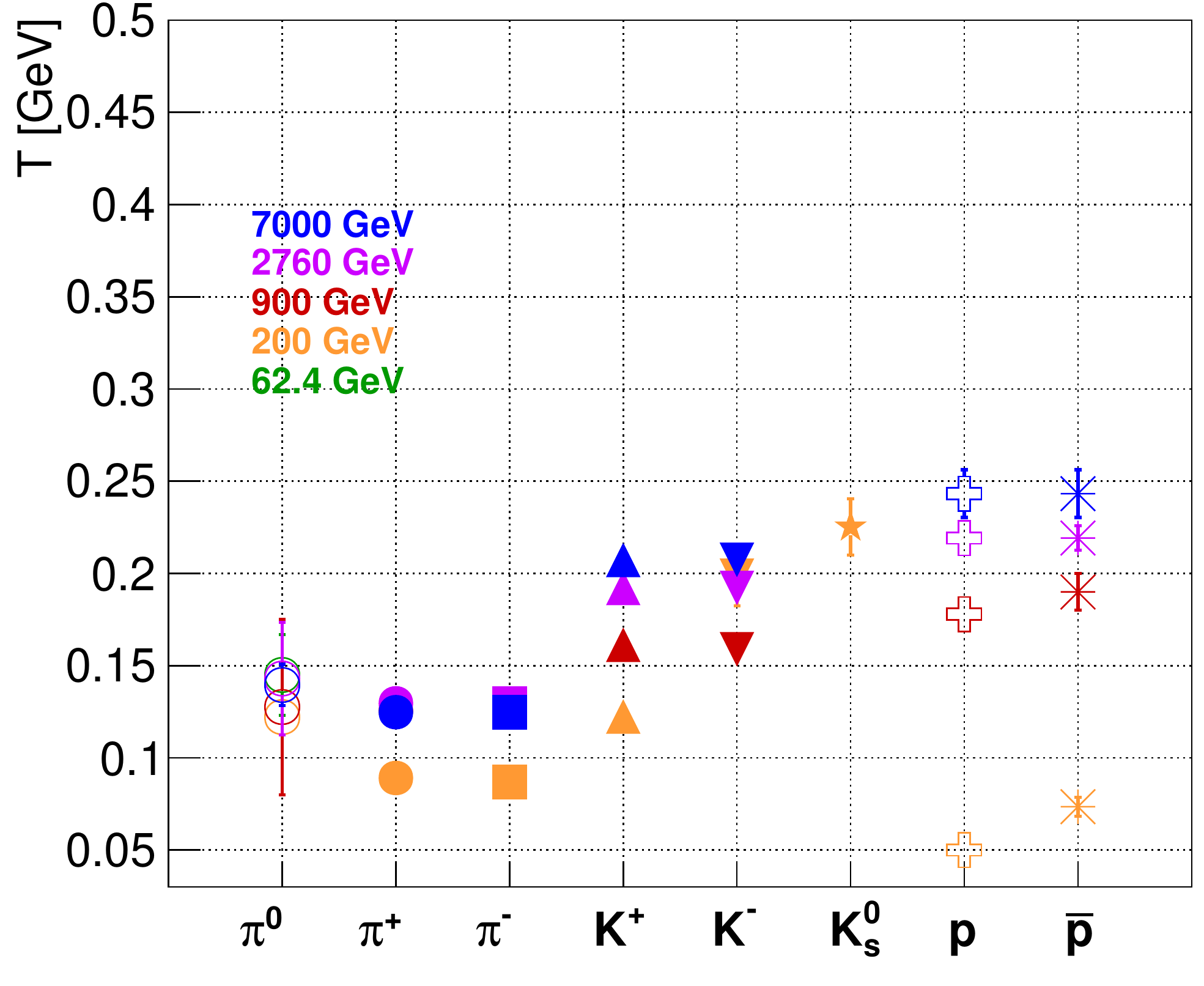}
	    \caption{}
    \end{minipage}
    \caption{The fitted $q_i$ and $T_i$ values for each hadron type, $i \in $ $\pi^{\pm}$, $\pi^{0}$, $K^{\pm}$, $K_s^0$, $p$, and $\bar{p}$, c.f.~\cite{artic:cleymansqspecies}.}
    \label{fig:qspecies}
\end{figure}

On Fig.~\ref{fig:qspecies} we show the fitted $q_i$ and $T_i$ for different hadron species at the above energies, $\sqrt{s}$. In agreement with Ref.~\cite{artic:cleymansqspecies}, we see that the non-extensivity parameter $q_i$ is less sensitive to the hadron mass, however the importance of center-of-mass energy of the colliding system is remarkable. In contrast to this, parameter $T_i$ reflects the mass-hierarchy ordering.

\section{The $T-q$ PARAMETER SPACE FOR IDENTIFIED HADRONS}

Summarizing, the center-of-mass energy evolution of the fit parameters works well with our double-log assumptions. The $q_{2i}$ and $T_{2i}$ parameters are getting slightly larger with the larger hadron mass, $m_i$ and applying forlmulae~(\ref{eq:q-evol}) and~(\ref{eq:T-evol}) the evolution is described nicely in the whole tested energy range, $62.4$ GeV $\leq \sqrt{s} \leq 7$ TeV. 

The obtained $q_i(\sqrt{s})$ function increases with $\sqrt{s}$ in the range $1.07-1.17$ representing the deviation from the Boltzmann\,--\,Gibbs case with $q=1$.  The deviance from this case $q$ grows with higher center-of-mass energy values. The temperature-like parameters $T_i(\sqrt{s})$ show almost constant values, with the following $m_i$ hadron mass hierarchy: $T_{\pi}=120-140$ MeV, $T_{K}=120-200$ MeV, and $T_p=70-240$ MeV. 

\begin{figure}[!h]
  \centerline{\includegraphics[width=0.65\textwidth]{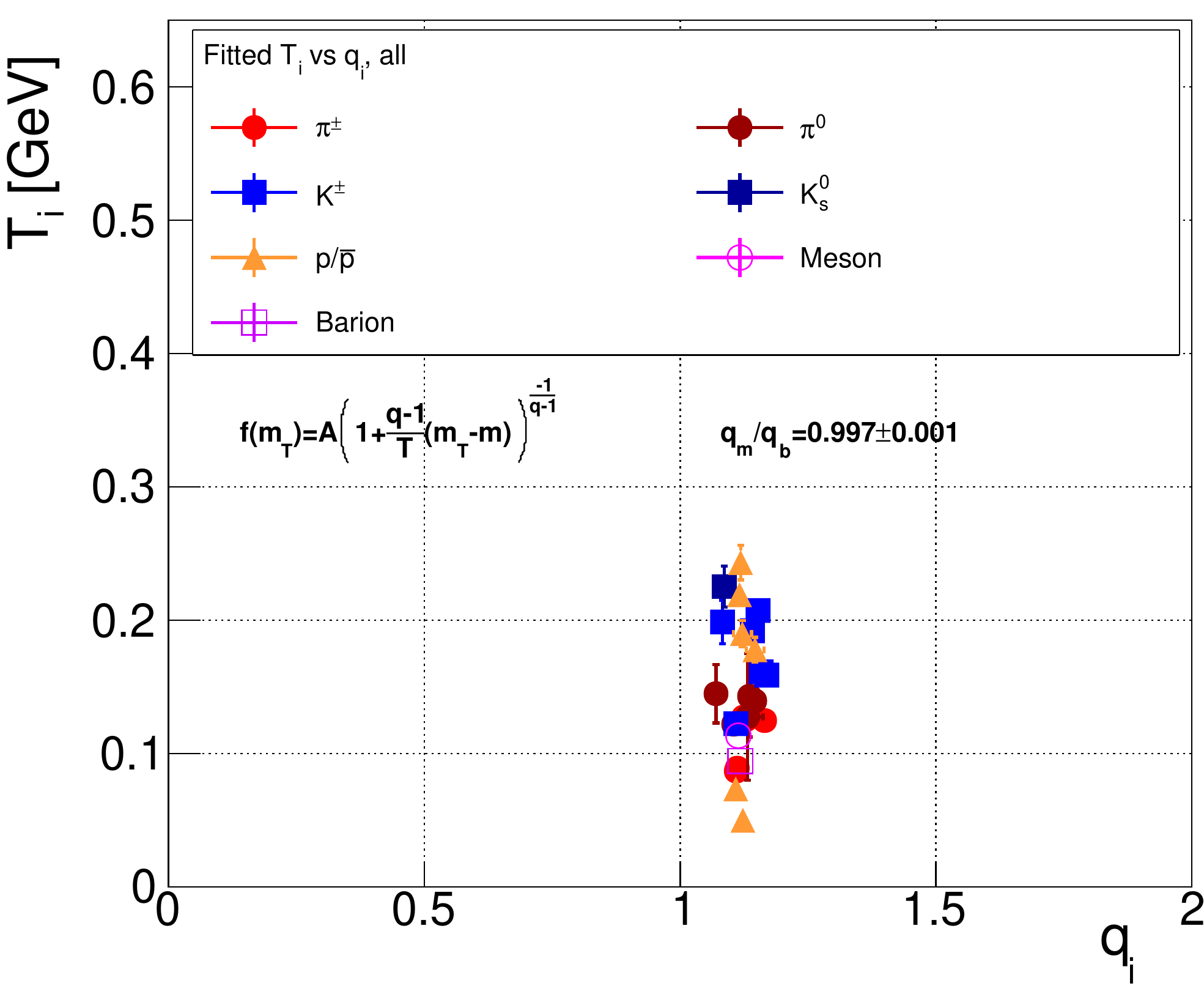}}
  \caption{The parameter space $T_i-q_i$ for hadron type, $i \in $ $\pi^{\pm}$, $\pi^{0}$, $K^{\pm}$, $K_s^0$, $p$, and $\bar{p}$.}
  \label{fig:T-q}
\end{figure}

We plotted the parameters $q_i$ and $T_i$ in Fig.~\ref{fig:T-q}. Points fitted separately at each $\sqrt{s}$ value, group nicely in the compact area $T_{i}=70-240$ MeV and $q_i(\sqrt{s})=1.07-1.17$. This suggests the validity of the non-extensive statistical approach presented in Ref.~\cite{artic:tsbeurphys2}. 

Beside the points of the hadron species, the averaged baryonic and mesonic point are also indicated in Fig.~\ref{fig:T-q}. The ratio is $q_m/q_b=0.997\pm 0.001$ as indicated on the plot, which presents the overlap.

%

\clearpage
\section{SUMMARY}

In this study we analyzed identified hadron spectra measured in proton-proton collisions from RHIC to LHC energies in the range $62.4$ GeV $\leq \sqrt{s} \leq 7$ TeV. We showed that the Tsallis\,--\,Pareto distributions originated from non-extensive thermodynamics describe the spectra very well in wide $m_T$ regions, typically $p_T\lesssim 10-20$ GeV/$c$ using the distribution in the form of~(\ref{eq:TS}). The $\sim \log(\log(\sqrt{s}))$-like evolution of the parameters $q_i$ and $T_i$ were tested on the identified hadron spectra data measured for $\pi^{\pm}$, $\pi^{0}$, $K^{\pm}$, $K_s^0$, $p$, and $\bar{p}$. It seems, both the non-extensivity, $q_i$ and temperature-like $T_i$ parameters agree with the suggested QCD inspired evolution pattern. However the temperature has almost constant value within the investigated $\sqrt{s}$ regime. We found a mass-ordered hierarchy in the evolution parameters, i.e. heavier hadrons have larger $q_{i}$ and $T_i$ values.
We found averaged mesonic and baryonic non-extensitivity parameters close to each other, namely $q_m/q_b \approx 1$.

\section{ACKNOWLEDGMENTS}
This work was supported by Hungarian OTKA grants K104260, NK106119, K120660, MTA-UA bilateral mobility program NKM-81/2016 and NIH TET 12 CN-1-2012-0016. Author GGB also thanks the J\'anos Bolyai Research Scholarship of the Hungarian Academy of Sciences. Author GB thanks for the support of Wigner GPU Laboratory.



\bibliographystyle{aipnum-cp}%
\bibliography{BIRO-me16-paper.bib}%

\end{document}